\documentclass[12pt]{article}
\usepackage{graphicx, here, amsfonts, epsfig, hhline}

\begin{document}

\begin{titlepage}
\hfill \vbox{\hbox{DFCAL-TH 03/3}
\hbox{February 2003}}

\vskip 0.5cm

\centerline{\bf THE POMERON IN EXCLUSIVE VECTOR MESON PRODUCTION}

\vskip 0.3cm

\centerline{R.~Fiore$^{a\dagger}$, L.L.~Jenkovszky$^{b\ddagger}$,
F.~Paccanoni$^{c\ast}$, A.~Prokudin$^{d\diamond}$}

\vskip 0.1cm

\centerline{$^{a}$ \sl Dipartimento di Fisica, Universit\`a della Calabria}
\centerline{\sl Instituto Nazionale di Fisica Nucleare, Gruppo collegato di Cosenza}
\centerline{\sl I-87036 Arcavata di Rende, Cosenza, Italy}

\centerline{$^{b}$ \sl Bogolyubov Institute for Theoretical Physics}
\centerline{\sl Academy of Science of Ukraine}
\centerline{\sl UA-03143 Kiev, Ukraine}

\centerline{$^{c}$ \sl Dipartimento di Fisica, Universit\`a di Padova}
\centerline{\sl Instituto Nazionale di Fisica Nucleare, Sezione di Padova}
\centerline{\sl Via F. Marzolo 8, I-35131 Padova, Italy}

\centerline{$^{d}$ \sl  Dipartimento di Fisica Teorica, Universit\`a
Degli Studi di Torino}
\centerline{\sl Instituto Nazionale di Fisica Nucleare, Sezione di Torino}
\centerline{\sl Via P. Giuria 1, I-10125 Torino, Italy}
\centerline{\sl Institute For High Energy Physics, 142284 Protvino, Russia} 

\vskip 0.1cm
\begin{abstract}
An earlier developed model for vector meson photoproduction,
based on a dipole Pomeron exchange, is extended to electroproduction.
Universality of the non linear Pomeron trajectory is tested by fitting
the model to ZEUS and H1 data as well as to CDF data on $\bar pp$ elastic
scattering.

\end{abstract}

\vskip 0.1cm

\vfill

\hrule
$
\begin{array}{ll}
^{\dagger}\mbox{{\it e-mail address:}} &
  \mbox{FIORE@CS.INFN.IT} \\
^{\ddagger}\mbox{{\it e-mail address:}} &
  \mbox{JENK@GLUK.ORG} \\
^{\ast}\mbox{{\it e-mail address:}} &
  \mbox{PACCANONI@PD.INFN.IT} \\
^{\diamond}\mbox{{\it e-mail address:}} &
  \mbox{PROKUDIN@TO.INFN.IT} 
\end{array}
$
\end{titlepage}
\eject
\newpage

\section{Introduction}

Elastic production of vector mesons in electron-proton interaction
has provided a deeper understanding of the diffraction phenomenon
and finds a sensible description in a variety of models. The first
attempts to describe diffractive photoproduction were based on
Regge theory~\cite{DL} and vector dominance model~\cite{SC}. Since
various aspects of the deep inelastic scattering and of elastic
processes are both present in photoproduction it is quite
natural that perturbative QCD can help to understand many features
of the HERA experimental results. Examples where perturbative QCD
has been applied to this process can be found in 
Ref.~\cite{RY}. 

In these perturbative
calculations regarding diffractive processes, that in this case
have characteristic features of the elastic ones, non perturbative
contributions are present and their description becomes an
important ingredient of the theoretical model while lying outside
perturbation theory. Hence, models based on Regge pole
phenomenology maintain their important task in helping to
construct a representation of non perturbative aspects of the
scattering amplitude. For the processes under consideration many
papers based on Regge poles exchange successfully
reproduced~\cite{JMP,FJP} new experimental HERA data. 
A different approach to asymptotic behavior was advocated 
recently in Ref.~\cite{TT}.

The aim of this paper is to expound the properties of the
most important and intricate Regge exchange: the vacuum, or
Pomeron exchange. $J/\psi$ photoproduction and, apart from small
subleading contributions, $\Phi$ photoproduction are genuine
Pomeron filters that, together with very high energies reactions,
permit a careful study of the non perturbative features of
diffraction. 

After a short description of the model,
that has been already introduced in Ref.~\cite{FJP}, in Section 2,
an analysis of the HERA data for $J/\psi$ and $\Phi$ photoproduction
 and electroproduction will be presented in Sections 3 and 4.
 A test of the Pomeron universality where the
model is applied to proton-antiproton elastic scattering at very
high energy will be considered in Section 5.
The Conclusions in Section 6 will complete the paper.

\section{The model}

A convenient way to obtain rising cross sections with the Pomeron
intercept equal to one assumes that the Pomeron is a double Regge
pole. This means that the $t$-channel partial wave, corresponding to
the Pomeron exchange, has a double pole for $\ell=\alpha(t)$. In
this choice we are comforted by the numerous successes of this
model in its applications to hadronic reactions~\cite{JMP,FJP,LJ}.
The total cross section will then increase logarithmically with $s$
at large energies. 

The reason why non asymptotic
contributions are more important for the dipole than for the Regge
pole can be easily seen. The asymptotic behavior of
$P_{\alpha}(z)$, where $z=-1-2s/(t-4m^2)$, is $P_{\alpha}(z)\sim
z^{\alpha}(1+O(1/z^2))$. In the case of a dipole, the corrections
are of the order $1/z$ since a simple calculation gives, in the
limit $\alpha \to 1$ and $z\to \infty$,
\begin{displaymath}
\left. \frac{\partial P_\alpha(z)}{\partial \alpha} \right|_{\alpha=1}\sim z \left[\ln z-
4\pi+\frac{8\pi}{z}+O\left(\frac{1}{z^2}\right) \right].
\end{displaymath}
To this correction, that could influence the low energy behavior
of the amplitude, we must add the contribution of a Pomeron
daughter, that is difficult to estimate, and the possibility that
also quarks be present in the Pomeron. All these arguments suggest
that it is better to apply the model as far as possible from the
low energy region, choosing the invariant scattering amplitude in
the form
\begin{equation}
A(s,t)=i f(t)
\left(-i\frac{s}{s_0}\right)^{\alpha_P(t)}\left[\ln\left(-i\frac{s}{s_0}
\right)-g(t)\right]\;, \label{d1}
\end{equation}
where $g(t)$ and $f(t)$ are functions, for the moment
undetermined, of the momentum transfer. $\alpha_P(t)$ is the
Pomeron trajectory with $\alpha_P(0)=1$.

The choice
of the function $f(t)$, that represents the product of the
vertices $\gamma$-Pomeron-meson and proton-Pomeron-proton,
 will be made by
imposing the condition that the Pomeron exchange is pure spin
$\alpha_P$ exchange. It has been shown in~\cite{FP} that this
constraint leads to a vertex of the form $[(\alpha_P(t)-1)f_1(t)
+(\alpha_P(t)+1)f_2(t)]$ and, in the neighborhood of $t=0$, to a
term vanishing with $t$ whatever the form of the trajectory could
be. The presence of this term, that is usually neglected, makes
the determination of the slope independent of the value of the
differential cross section in the forward direction and can help
in explaining non trivial properties of the forward cone.

 It has been shown in Ref.~\cite{FJP} that the simple form for the
elastic differential cross section of vector meson photoproduction
\begin{equation}
\frac{d\sigma}{dt}=\left[a\,e^{bt}+ct\,e^{dt}\right]^2\left(
\frac{s}{s_0}\right)^{2\alpha_P(t)-2}\left[\left(\ln\frac{s}{s_0}+g\right)^2+
\frac{\pi^2}{4}\right]\;, \label{d2}
\end{equation}
where $g$ is a constant, gives a good quality fit to the
experimental data~\cite{ZEUS,H1}. We notice that the form
(\ref{d2}) satisfies also the aforesaid conditions and will be
adopted in this paper. 

Since the amplitude, in the Regge
form, should have no essential singularity at infinity in the cut
plane, ${\cal R}e\:\alpha(s)$ is bounded by a constant, for $s\to
\infty$, and this leads to the bound
\begin{displaymath}
|\alpha(s)|< Ms^q,\;\;\;\;\mbox{for}\;s\to\infty
\end{displaymath}
with $q<1$ and $M$ an arbitrary constant. The
choice~\cite{RF,FJP}
\begin{equation}
\alpha_P(t)=1+\gamma(\sqrt{t_0}-\sqrt{t_0-t}), \label{d3}
\end{equation}
where $t_0=4m_{\pi}^2$ and $\gamma=m_{\pi}/1\,GeV^2$, satisfies the
above conditions and reproduces the standard Pomeron slope at
$t=0$, $\alpha'_P(0)\simeq 0.25\;GeV^{-2}$. Eq.~(\ref{d3}) for the
trajectory defines uniquely the model for photoproduction. 

Consider now electroproduction of a vector meson. As
noticed in Refs.~\cite{ZEUS1,H11} a commonly adopted form for the $Q^2$
dependence of the $J/\Psi$ cross section is
\begin{equation}
\sigma_{tot}^{\gamma^*\,p\;\to J/\Psi\,p} \propto
\frac{1}{(1+Q^2/M^2_{J/ \Psi})^n}\;\; , \label{d4}
\end{equation}
where $n\sim 1.75$, according to the ZEUS Collaboration~\cite{ZEUS1}, and $n\sim
2.38$ according to the H1 Collaboration~\cite{H1}. 

For large $Q^2$ all the
amplitudes but the double flip one, for diffractive vector meson
electroproduction, can be evaluated in perturbative QCD~\cite{
KNZ}. In the longitudinal photon amplitudes, a factor
$Q/M_{J/\Psi}$ is a consequence of gauge invariance irrespective
of the detailed production dynamics. If we consider only the
dominant twist $s$-channel helicity conserving amplitudes, the
factor in Eq.~(\ref{d4}) thus finds a natural explanation. The
$Q^2$ dependence, however, will appear also in the strong coupling
and in the gluon structure function through the hard scale of
perturbative QCD~\cite{KNZ}. 

In our
approach, based on Regge pole theory, the factor (\ref{d4}) will
be certainly present in electroproduction, multiplying the
differential cross section (\ref{d2}), but this will not complete
all the possible corrections. Since, in the dipole Pomeron
formalism, the product of the vertices can affect the parameter
$g$, all the parameters can acquire a weak $Q^2$ dependence. We
neglect this dependence in $a,\,b,\,c,\,d $ and assume that $g$
varies as $g\times [1+Q^2/(Q^2+M_V^2)]^{\gamma}$ where $\gamma$,
if this assumption is correct, is small. One can interpret this
functional dependence of $g$ as coming from a $Q^2$ dependence of
$s_0$ in $\ln(s/s_0)$.  \vskip 0.3cm The final form of the
differential cross section is:
\begin{displaymath}
\frac{d\sigma}{dt}= \left (1+\frac{Q^2}{M^2_{J/\Psi}}
\right)^{-\beta} \left [ a\,e^{bt}+ct\,e^{dt}\right]^2
\left(\frac{s}{s_0} \right)^{2\alpha_P (t)-2}\times
\end{displaymath}
\begin{equation}
\left(\left[\ln\left(\frac{s}{s_0}\right)+g\;[1+Q^2/(Q^2+M_V^2)]^{\gamma}\right]^2+
\frac{\pi^2}{4}\right)\;, \label{d5}
\end{equation}
where, for $Q^2=0$, all the parameters have the same value as
for photoproduction. We notice that the value of $\beta$ includes a factor
$(1+Q^2/M^2_{J/\Psi})$ that comes from the contribution of the longitudinal
 amplitude, relevant at $Q^2\neq 0$, which leads to
$|A|^2=|A_T|^2+|A_L|^2$. The approximate relation 
$A_L \sim Q\;A_T/M_{J/\Psi}$ can be applied in this phenomenological approach. 

In the following Sections
the parameterizations (\ref{d2}) and (\ref{d5}) will be applied to $J/\Psi$
and $\Phi$ photoproduction and electroproduction. Only for these
processes, in fact, the dominant exchange is due to the Pomeron
and we do not need to introduce non-leading trajectories. We can
thus avoid contributions that can interfere with the object of our
study and that, at intermediate energies where data are available,
can be large.

\section{$J/\Psi$ photoproduction and electroproduction}
Following the analysis of Ref.~\cite{FJP} we apply Eq.~(\ref{d2}) to the
new dataset of $J/\Psi$ photoproduction \cite{ZEUS}\footnote{The data are available from \cite{data}.}. This dataset represents measurements of $J/\Psi$ photoproduction
in two channels: $J/\Psi\rightarrow e^+e^-$ and
 $J/\Psi\rightarrow \mu^+\mu^-$ which may possibly have different
normalizations especially for differential cross sections. To test
if it is so we limit our fit to the region $W\le 160\; {\rm GeV}$ and check the predictions
of the model for the differential cross section and the total integrated
cross section.

As noticed in the previous paper \cite{FJP}, the parameter $d$ varies little 
in the fit, so that we keep the same value $d=0.851\; {\rm GeV^{-2}}$  
fixed, thus leaving only four parameters free. In order to avoid the
region of inelastic background we limit the $t$ region to $|t|<1\; {\rm  GeV^2}$.
In the fit we use differential cross sections only.
For the electron channel we have obtained the results shown
in Column 2 of Table~\ref{Table 1.}, with $\chi^2/{\rm d.o.f.}=1.5$.
For the muon channel the results are presented in Column 3 of 
Table~\ref{Table 1.}, with $\chi^2/{\rm d.o.f.}=1.0$.

{\small
\begin{table}[H]
\begin{center}
\begin{tabular}{|l|l|l|l|}
\hline
\multicolumn{4}{c}{Photoproduction} \\
\hline
 & $J/\Psi \rightarrow e^+e^-$ & 
   $J/\Psi \rightarrow \mu^+\mu^-$ & 
   $J/\Psi \rightarrow e^+e^-$   
   \\
 & $W<160\; {\rm GeV}$ & 
   $W<160\; {\rm GeV}$ & 
   $W<300\; {\rm GeV}$  
   \\
\hline
a $[{\rm GeV}^{-1}]$ & 
  $(1.8 \pm 0.1)\cdot 10^{-3}$ &
  $(1.83 \pm 0.09)\cdot 10^{-3}$ &
  $(1.97 \pm 0.13)\cdot 10^{-3}$ 
  \\
b $[{\rm GeV}^{-2}]$ & 
  $1.55   \pm    0.49$ &
  $2.25   \pm    0.24$ &
  $1.40   \pm    0.51$ 
  \\
c $[{\rm GeV}^{-3}]$ & 
  $(-0.48 \pm 0.54)\cdot 10^{-3}$ &
  $(-0.97 \pm 0.19)\cdot 10^{-3}$ &
  $(-0.35 \pm 0.67)\cdot 10^{-3}$
 \\
d $[{\rm GeV}^{-2}]$ & 
  $0.851$ &
  $0.851$ &
  $0.851$  
  \\
g  & 
  $-4.23  \pm     0.37$ &
  $-4.25  \pm     0.22$ &
  $-4.58  \pm     0.29$ 
\\
\hline
%
\end{tabular}
\end{center}
\vskip -0.5cm \caption{ Values of parameters obtained by fitting
$J/\Psi$ photoproduction data.  
\label{Table 1.}}
\end{table}}

\begin{figure}[H]
\centering {
\epsfysize=90mm \epsffile{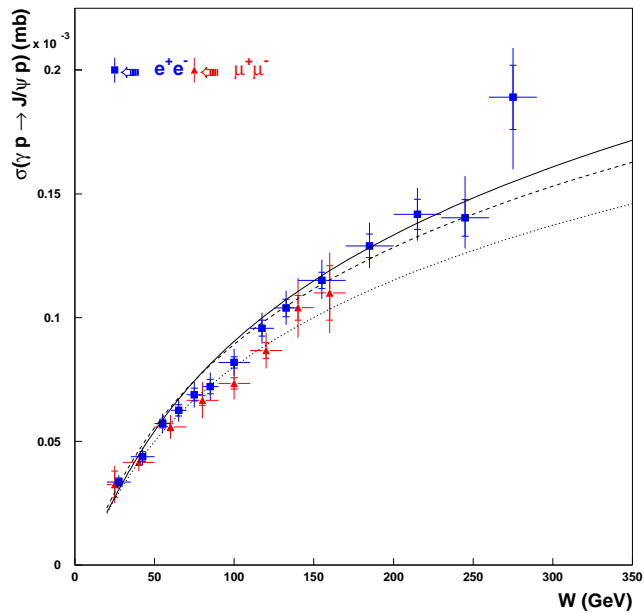}} 

\caption{Elastic
cross section of $J/\Psi$ photoproduction.
The dashed line corresponds to $J/\Psi \rightarrow e^+e^-$ channel fit
(Column 2 of Table \ref{Table 1.}). The dotted line corresponds to 
$J/\Psi \rightarrow \mu^+\mu^-$ channel fit
(Column 3 of Table \ref{Table 1.}). The solid line corresponds to 
$J/\Psi \rightarrow e^+e^-$ channel fit
(Column 4 of Table \ref{Table 1.}).
\label{fig:jpsi} }
\end{figure}

\begin{figure}[H]

\parbox[c]{7.cm}{\epsfxsize=80mm
\epsffile{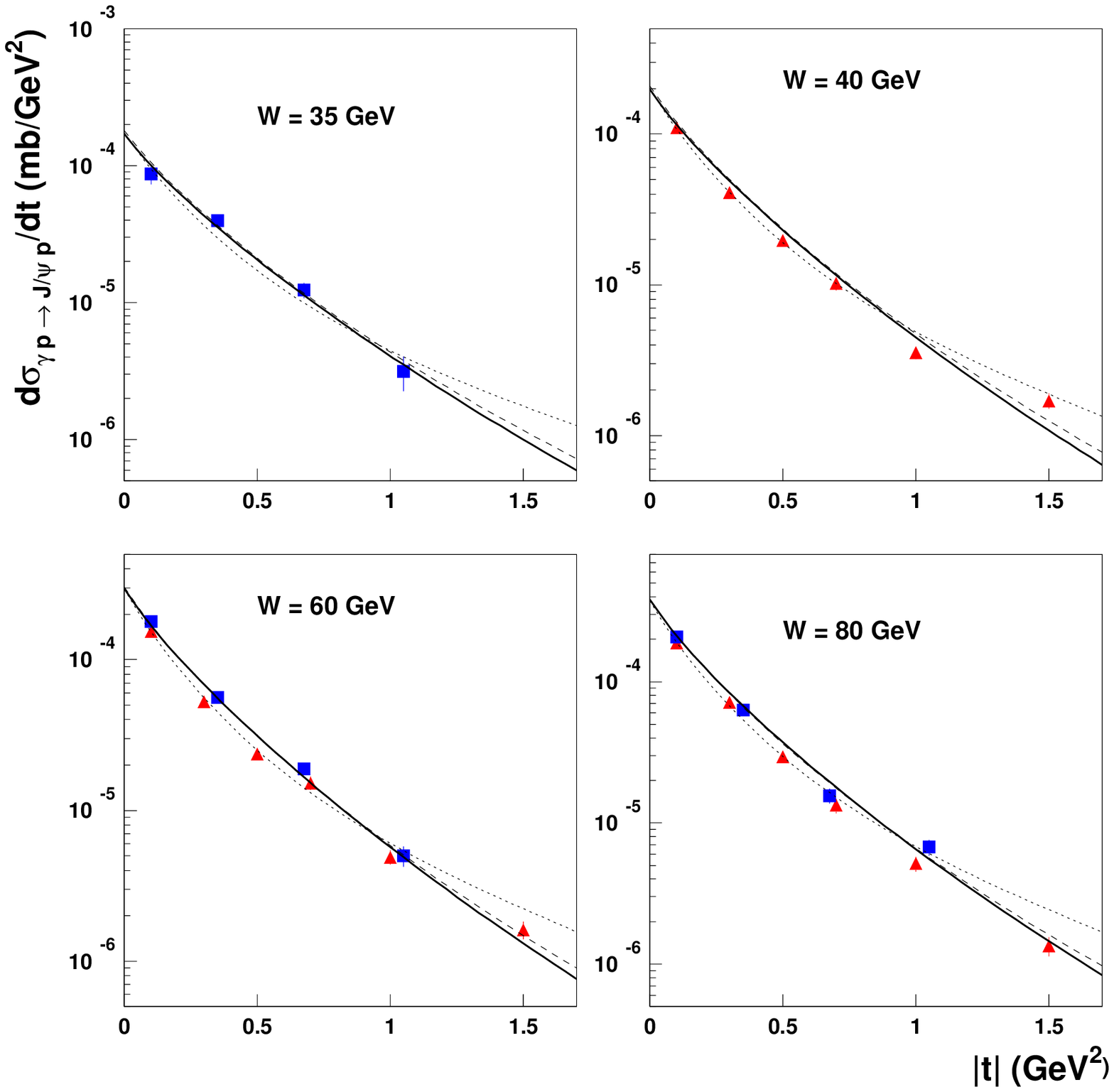}} \hfill~\parbox[c]{7.cm}{\epsfxsize=80mm
\epsffile{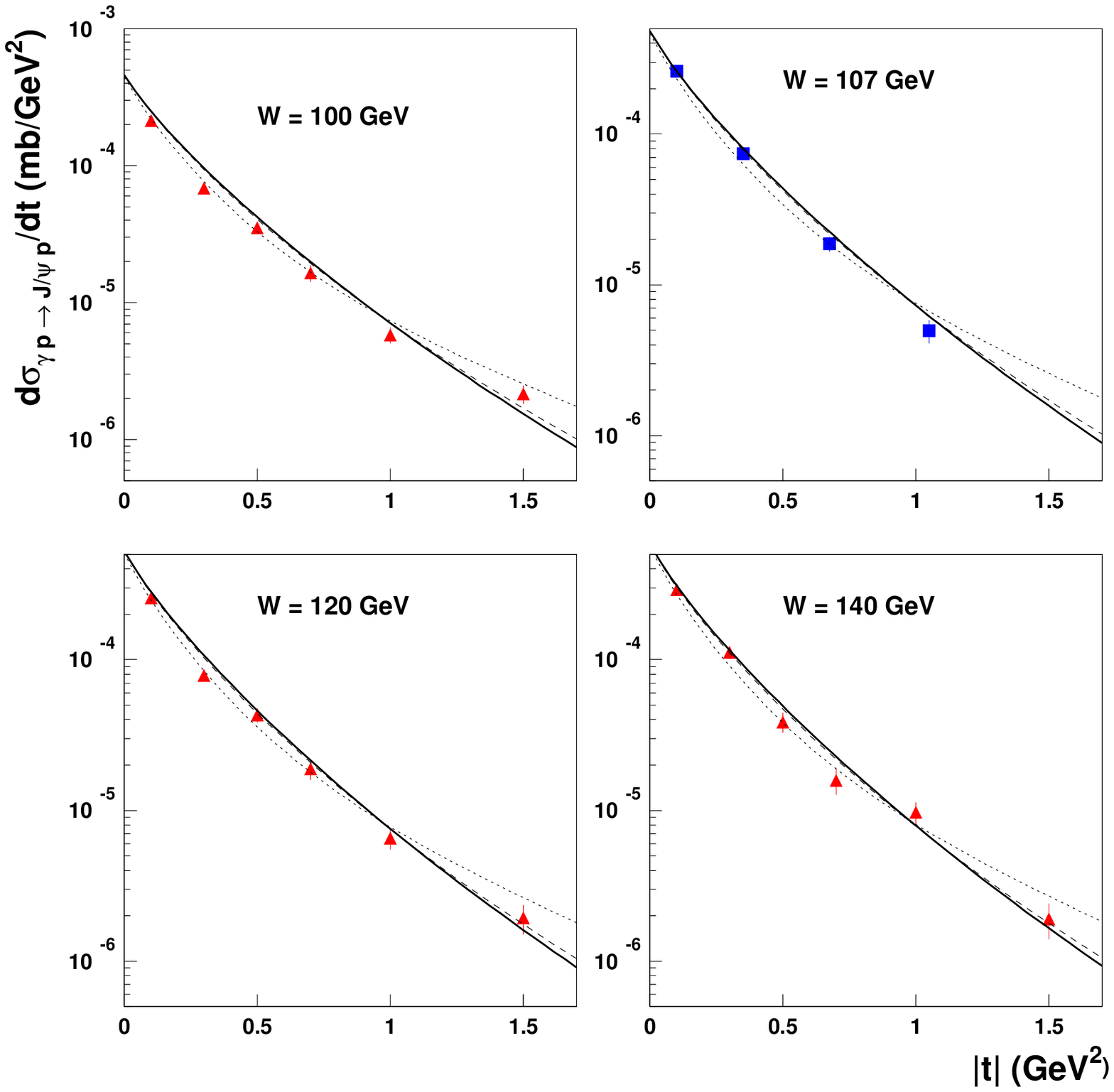}}

\parbox[t]{7.cm}{\caption{Differential cross section of
exclusive $J/\Psi$ photoproduction
for 35$\le W\le$ 80 GeV. Line aliases and symbols are the same as
in Fig. \ref{fig:jpsi}.
\label{fig:jpsid}}}
\hfill~\parbox[t]{7.cm}{\caption{Differential cross section of
exclusive $J/\Psi$ photoproduction
for 100$\le W\le$ 140 GeV. Line aliases and symbols are the same as
in Fig. \ref{fig:jpsi}.
\label{fig:jpsid1}}}
\end{figure}

\begin{figure}[H]
\parbox[c]{7.cm}{\epsfxsize=80mm
\epsffile{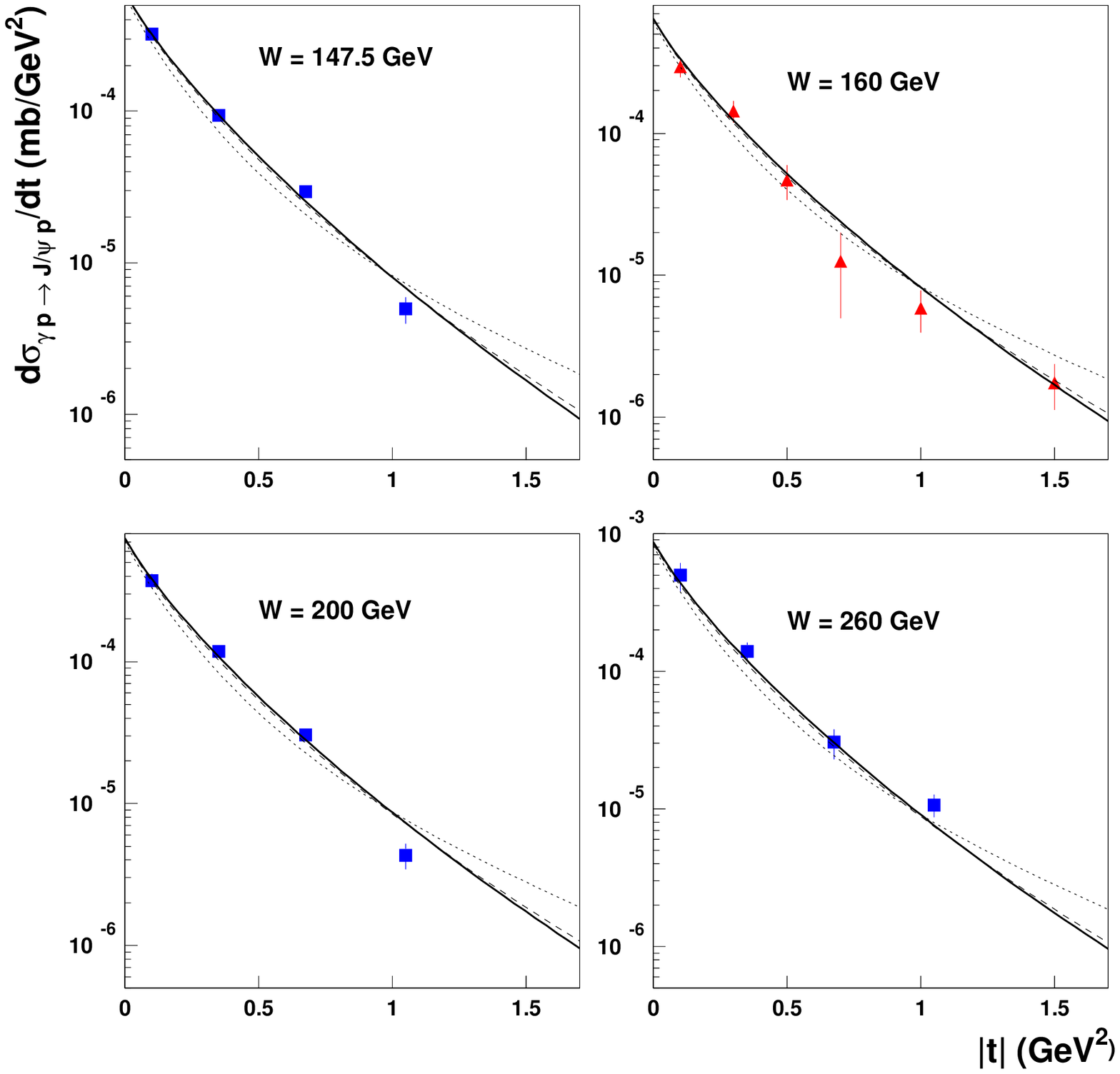}} \hfill~\parbox[c]{7.cm}{\epsfxsize=80mm
\epsffile{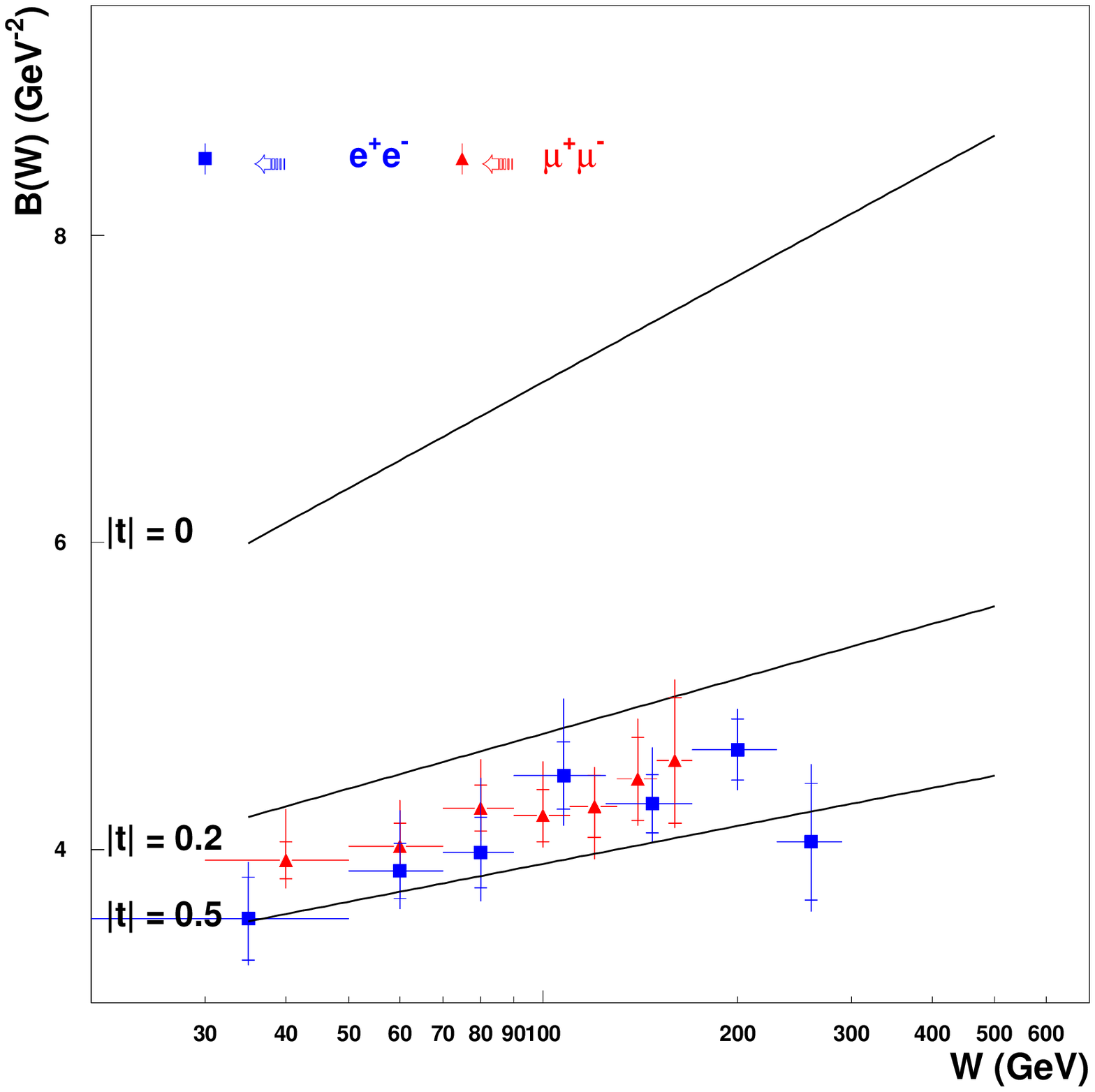}}

\parbox[t]{7.cm}{\caption{Differential cross section of
exclusive $J/\Psi$ meson photoproduction
for 147.5$\le W\le$ 260 GeV. Line aliases and symbols are the same as
in Fig. \ref{fig:jpsi}.
\label{fig:jpsid2}}}
\hfill~\parbox[t]{7.cm}{\caption{The slope of differential cross section of
exclusive $J/\Psi$ photoproduction.
\label{fig:slope}}}
\end{figure}

The different channels give us different values
of parameters $b$ and $c$ and as can be seen in 
Figs.~\ref{fig:jpsi},~\ref{fig:jpsid},~\ref{fig:jpsid1},~\ref{fig:jpsid2}
the resulting high energy predictions for elastic and differential cross sections
of exclusive $J/\Psi$ photoproduction
are quite different. We notice that the electron channel predictions 
are in a very good agreement with the data for
$W>160\; {\rm GeV}$, while the muon channel 
provides somewhat lower estimate of cross sections for higher
energies. 
It will be interesting to see if the experimental
data on $J/\Psi \rightarrow \mu^+\mu^-$ channel would support
such a trend or the conclusion is only a statistical fluctuation. 
If we use all the two channel data altogether we obtain a very
high $\chi^2/{\rm d.o.f.} = 1.9$. 
To implement a better analysis one needs a more complete set
of data on differential cross sections in both channels.

We proceed by choosing the electron channel without limiting
the value of $W$. The result is presented in Column 4 of 
Table~\ref{Table 1.}, with $\chi^2/{\rm d.o.f.}=1.2$.

The behavior of the local slope $B(W)=\frac{d}{dt}(\ln \frac{d\sigma}{dt})$
as a function of $W$ for various fixed values of $|t|$ is presented in
Fig.~\ref{fig:slope}. As expected from the differential cross section 
and the curvature of the Pomeron 
trajectory~(\ref{d3}), the local slope decreases with $|t|$. Its value at 
$t\approx -(0.2 \div 0.5)$ GeV$^2$ meets the experimental measurements
(see Fig.~\ref{fig:slope}). This is
quite understandable, since the experimental value is the average over a wide 
interval in $t$ covering the measurements. 
The rise of the slope toward $t=0$ is a 
well-known phenomenon in hadronic physics; its appearance in photoproduction was
emphasized e.g. in Ref.~\cite{Nem}.

Without any fitting we achieve a good agreement with the data
on integrated elastic cross section, $\chi^2/{\rm point} = 0.95$. 
The high error of $c$ is due to the scarcity of the data on
the differential cross section in the region $0<|t|<1\; {\rm GeV^2}$.
A more complete set of high accurate data will allow us
to arrive at a definite conclusion about the values
of parameters.  

Now we use Eq. (\ref{d5}) in order to describe electroproduction
of $J/\Psi$. We fix all the parameters obtained by fitting
the photoproduction data (see Column 4 of Table \ref{Table 1.}.) and
fit two parameters $\beta$ and $\gamma$ to the dataset \footnote{The data are available from \cite{data}, \cite{data1}.}.
The values of the parameters are the following:
$\beta = 1.94 \pm 0.42$, $\gamma = 0.69   \pm   0.24$ and
$\chi^2/{\rm d.o.f.} = 0.81 $.

The factor $[1+Q^2/(Q^2+M_V^2)]^{\gamma}$ grows up to
1.5 in the available region of photon virtuality $0<Q^2<50\; {\rm GeV^2}$.

In the case of $\gamma = 0$ we obtain $\beta=2.86   \pm    0.09$
and $\chi^2/{\rm d.o.f.}=1.07$. We proved that the fit is rather 
insensible to the value of $0<\gamma<1$. 

\begin{figure}[H]
\parbox[c]{7.cm}{\epsfxsize=80mm
\epsffile{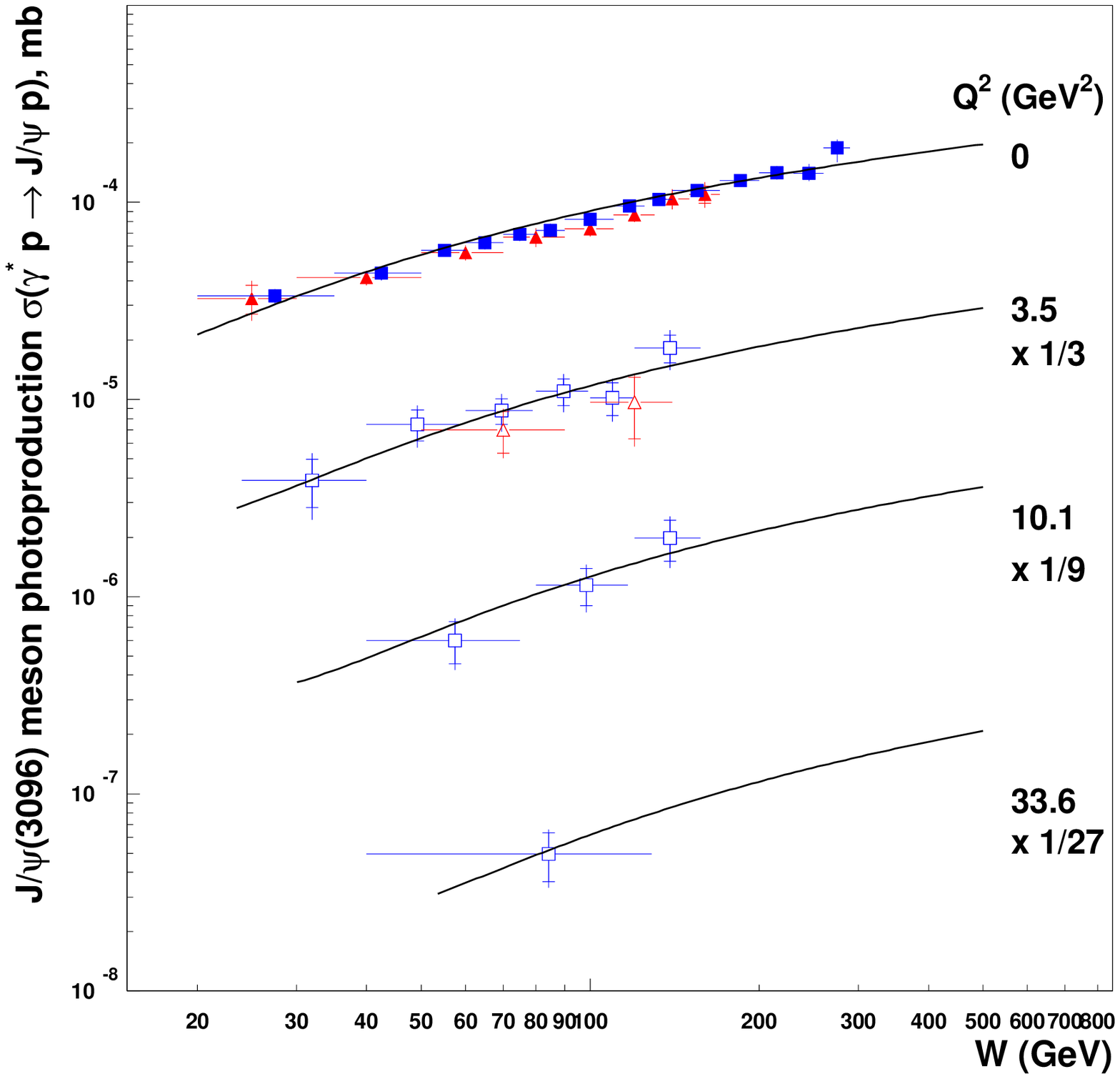}} \hfill~\parbox[c]{7.cm}{\epsfxsize=80mm
\epsffile{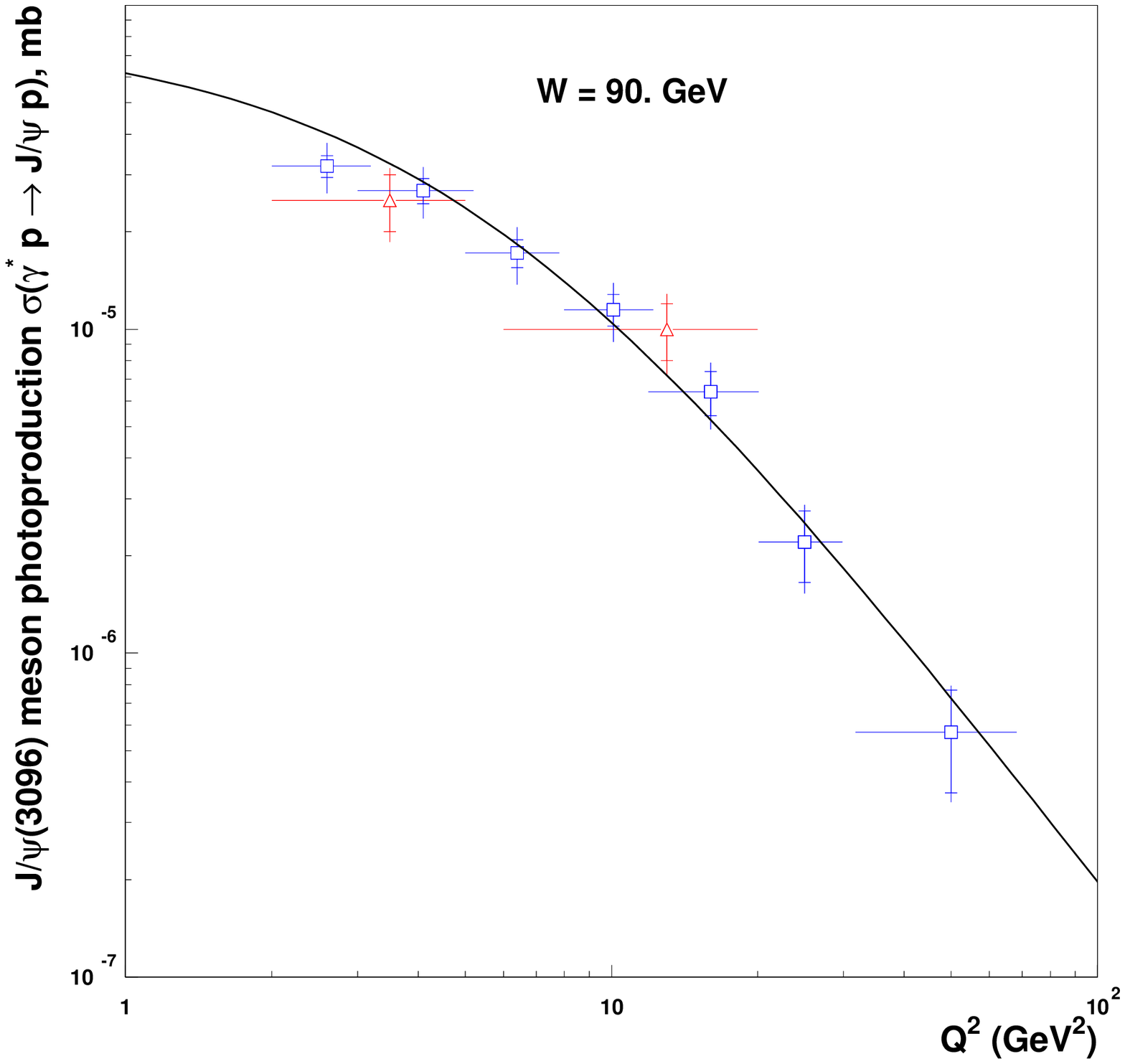}}

\parbox[t]{7.cm}{\caption{Cross section of
exclusive $J/\Psi$ electroproduction 
as a function of $W$. 
\label{fig:jpsiq}}}
\hfill~\parbox[t]{7.cm}{\caption{Cross section of
exclusive $J/\Psi$ electroproduction 
as a function of $Q^2$.
\label{fig:jpsiq1}}}
\end{figure}

\section{$\Phi$ meson photoproduction and electroproduction}

\begin{figure}[H]
\parbox[c]{7.cm}{\epsfxsize=80mm
\epsffile{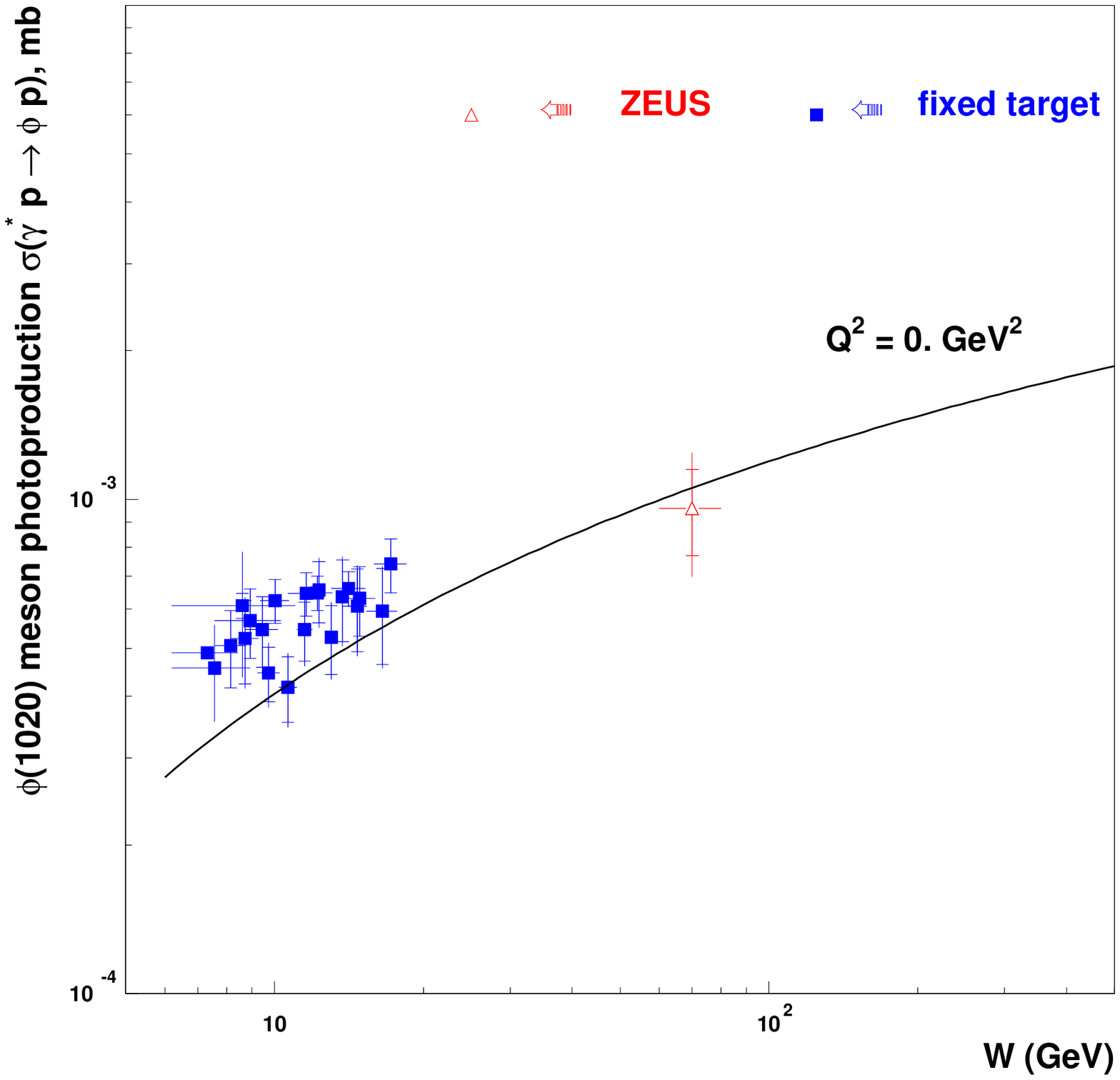}} \hfill~\parbox[c]{7.cm}{\epsfxsize=80mm
\epsffile{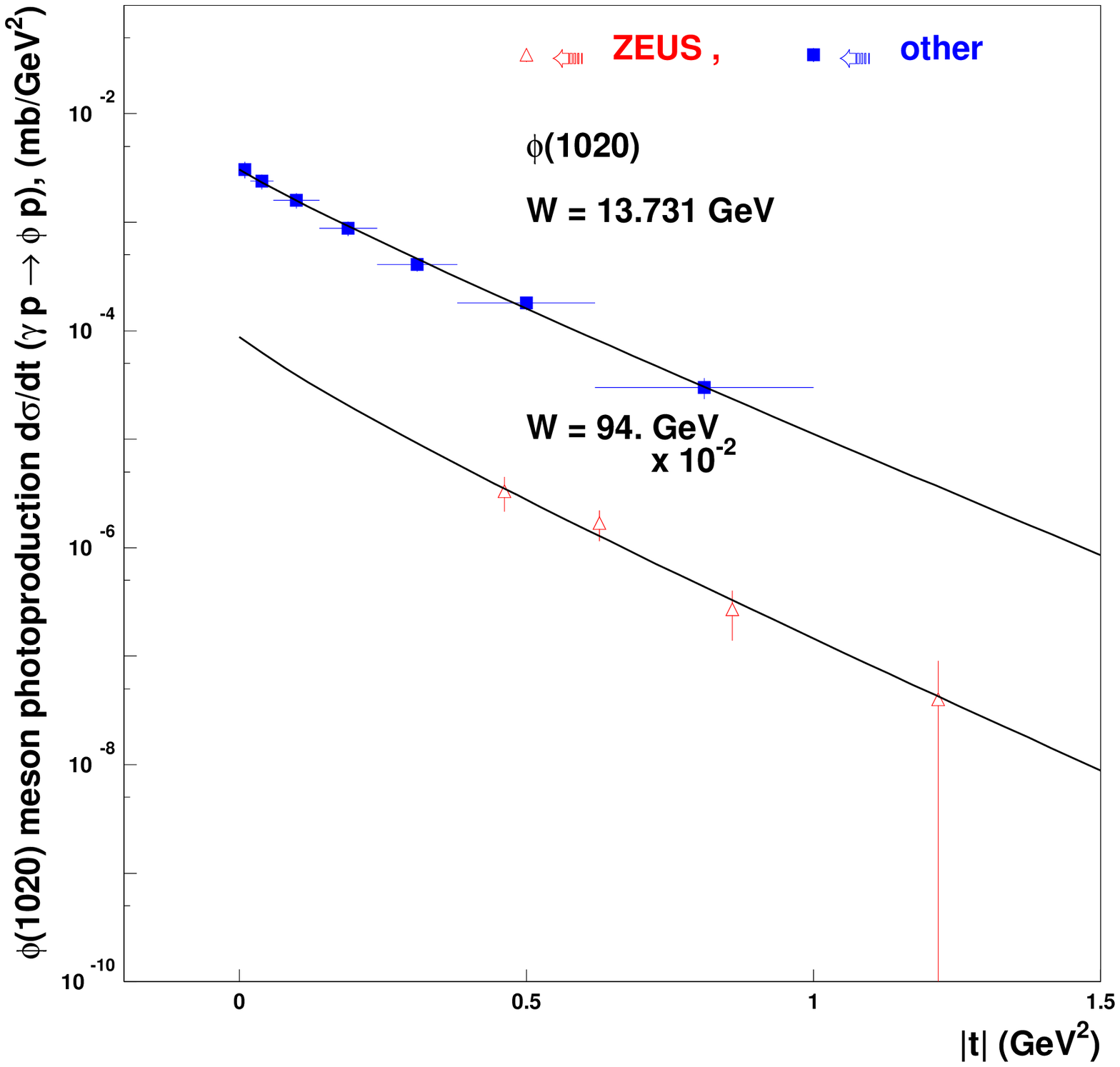}}

\parbox[t]{7.cm}{\caption{Elastic cross section of
exclusive $\Phi$ photoproduction. 
\label{fig:phi}}}
\hfill~\parbox[t]{7.cm}{\caption{Differential cross section of
exclusive $\Phi$ photoproduction.
\label{fig:phid}}}
\end{figure}

The major part of the data on $\Phi$ photoproduction
is concentrated in the low energy region$^2$, we thus
would not expect the asymptotic contribution (\ref{d2}) to work as 
well as in the case of $J/\Psi$ photoproduction. As we have only four
experimental points of the differential cross section at 
the energy $94$ GeV which are
in the region of intermediate $t$, we do not expect to obtain quantitative,
but rather qualitative description of the process (as can be seen in 
Figs.~\ref{fig:phi},~\ref{fig:phid} where exclusive elastic and
differential cross sections are presented). It turns out that the scarce 
data do not allow us to determine the parameters $c$ and 
$g$ with a reasonable error. The result of the fit is presented in 
Table~\ref{Table 2.} and the $\chi^2/{\rm d.o.f.} = 0.34 $.
{\small
\begin{table}[H]
\begin{center}
\begin{tabular}{ll}
a = & $(0.46 \pm 0.12)\cdot 10^{-2}\; [{\rm GeV}^{-1}]$, \\
b = & $2.26   \pm    0.12\; [{\rm GeV}^{-2}]$, \\
c = & $0.0 \pm 0.68\cdot 10^{-2}\;[{\rm GeV}^{-3}]$, \\
d = & $0.851\; [{\rm GeV}^{-2}]$, \\
g = & $-0.08  \pm 1.5$. \\
\end{tabular}
\end{center}
\vskip -0.5cm \caption{ Values of parameters obtained by fitting
$\Phi$ photoproduction data. (
$|t|<1\; {\rm GeV}^2$).  
\label{Table 2.}}
\end{table}}

As expected, the asymptotic formula (\ref{d2}) does not reproduce
the data near threshold (see Fig.~\ref{fig:phi}). 

\begin{figure}[H]
\parbox[c]{7.cm}{\epsfxsize=80mm
\epsffile{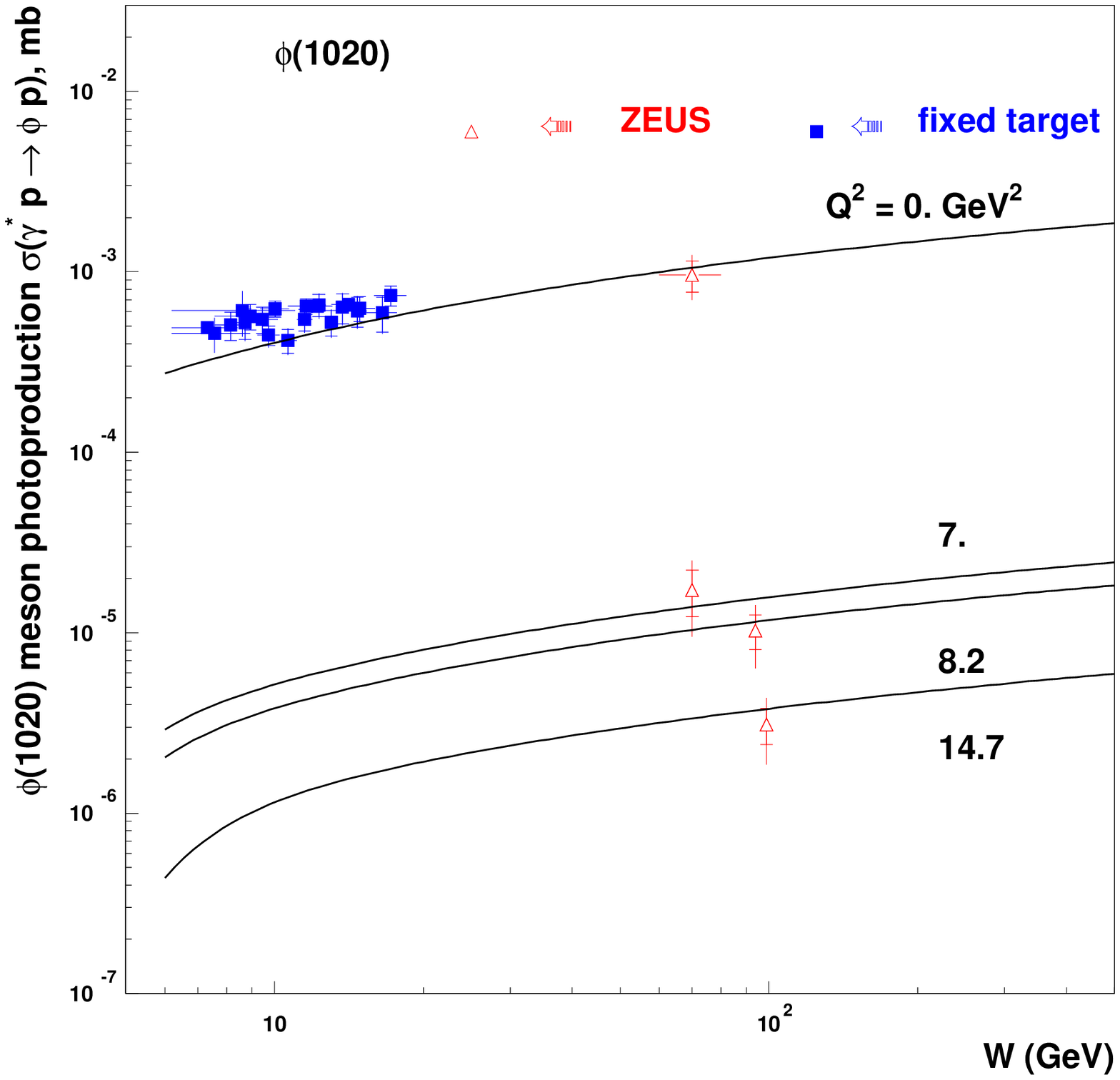}} \hfill~\parbox[c]{7.cm}{\epsfxsize=80mm
\epsffile{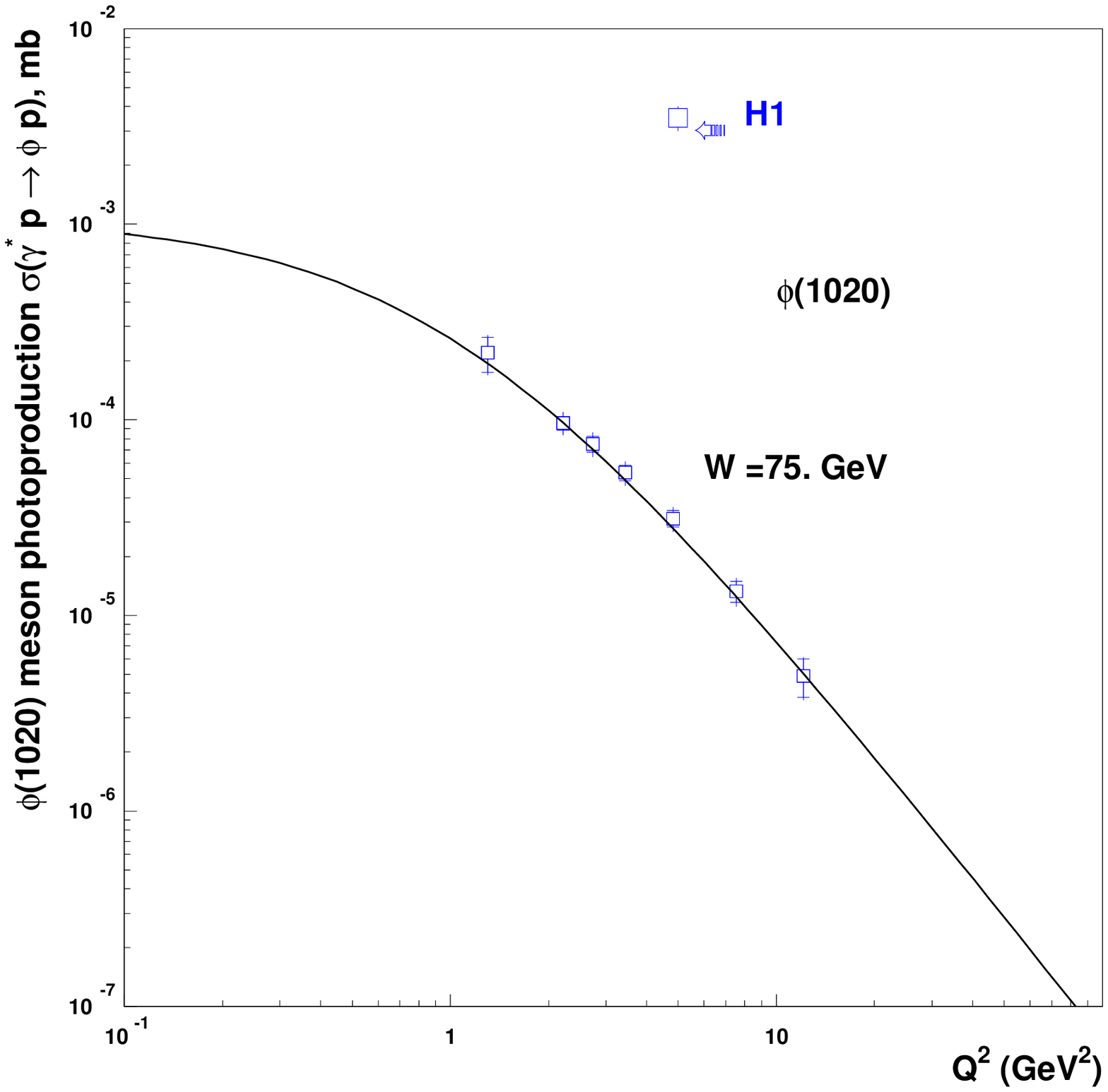}}

\parbox[t]{7.cm}{\caption{Elastic cross section of
exclusive $\Phi$ electroproduction as function of $W$. 
\label{fig:phiall}}}
\hfill~\parbox[t]{7.cm}{\caption{Elastic  cross section of
exclusive $\Phi$ electroproduction as function of $Q^2$.
\label{fig:phiq}}}
\end{figure}

To reproduce electroproduction of $\Phi$ we go along the
same lines as in the previous section and use  Eq. (\ref{d5}). 
Relying on our experience of $J/\Psi$ electroproduction,
 we restrict our
consideration to the case $\gamma = 0$ and obtain 
$\beta = 2.12\pm 0.03$, $\chi^2/{\rm d.o.f.}=0.3$. The results can
be seen
in Figs.~\ref{fig:phiall},~\ref{fig:phiq}, where elastic and 
differential cross sections of $\Phi$ exclusive electroproduction
are shown. 

\section{Pomeron universality}

The model we consider is consistent with $s$-channel unitarity and asymptotic
factorizability. Universality, in this context, refers to the choice
(\ref{d3}) for the Pomeron trajectory that provides a reliable description
of exclusive vector meson production. The conjecture that the trajectory
in Eq. (\ref{d3}) is universal is supported by the following example.

We consider the proton-antiproton scattering at sufficiently high energies,
where only the Pomeron presumably contributes. Following tradition \cite{DL1},
it is a customary practice to adopt a linear Pomeron trajectory in order 
to describe hadronic interactions. In a different approach \cite{LJ,JJS}
that provides a satisfactory fit to $pp$ and $\bar pp$ data a square root 
trajectory similar to  that of Eq.~(\ref{d3}) has been preferred. It is 
interesting to update this last fit using Eq.~(\ref{d3}) and the same 
parameters adopted for photoproduction: 
$t_0=4 m_\pi^2$ and $\gamma=m_\pi/1 {\rm GeV^2}$.

\begin{figure}[H]
\parbox[c]{7.cm}{\epsfxsize=80mm
\epsffile{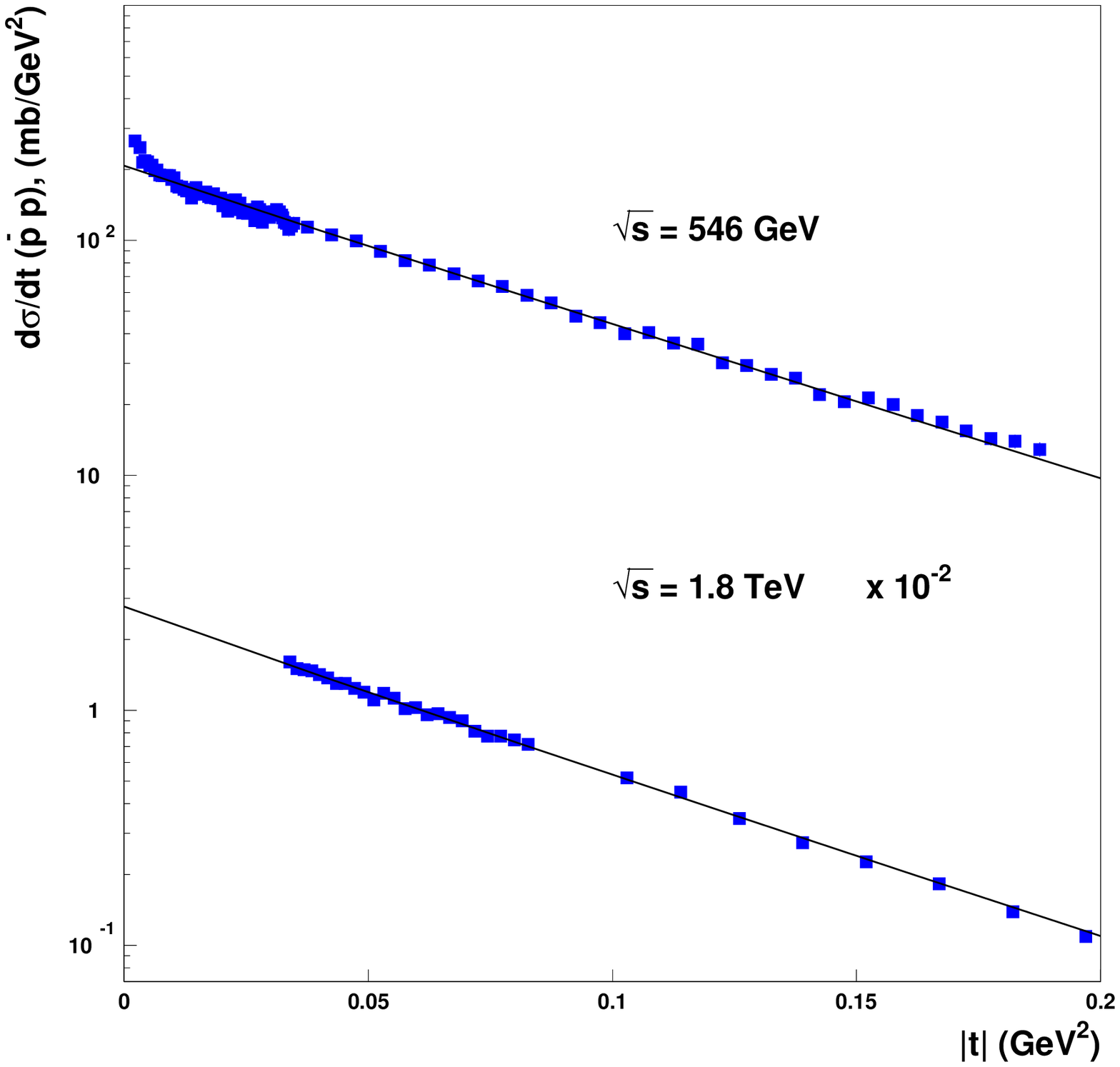}} \hfill~\parbox[c]{7.cm}{\epsfxsize=80mm
\epsffile{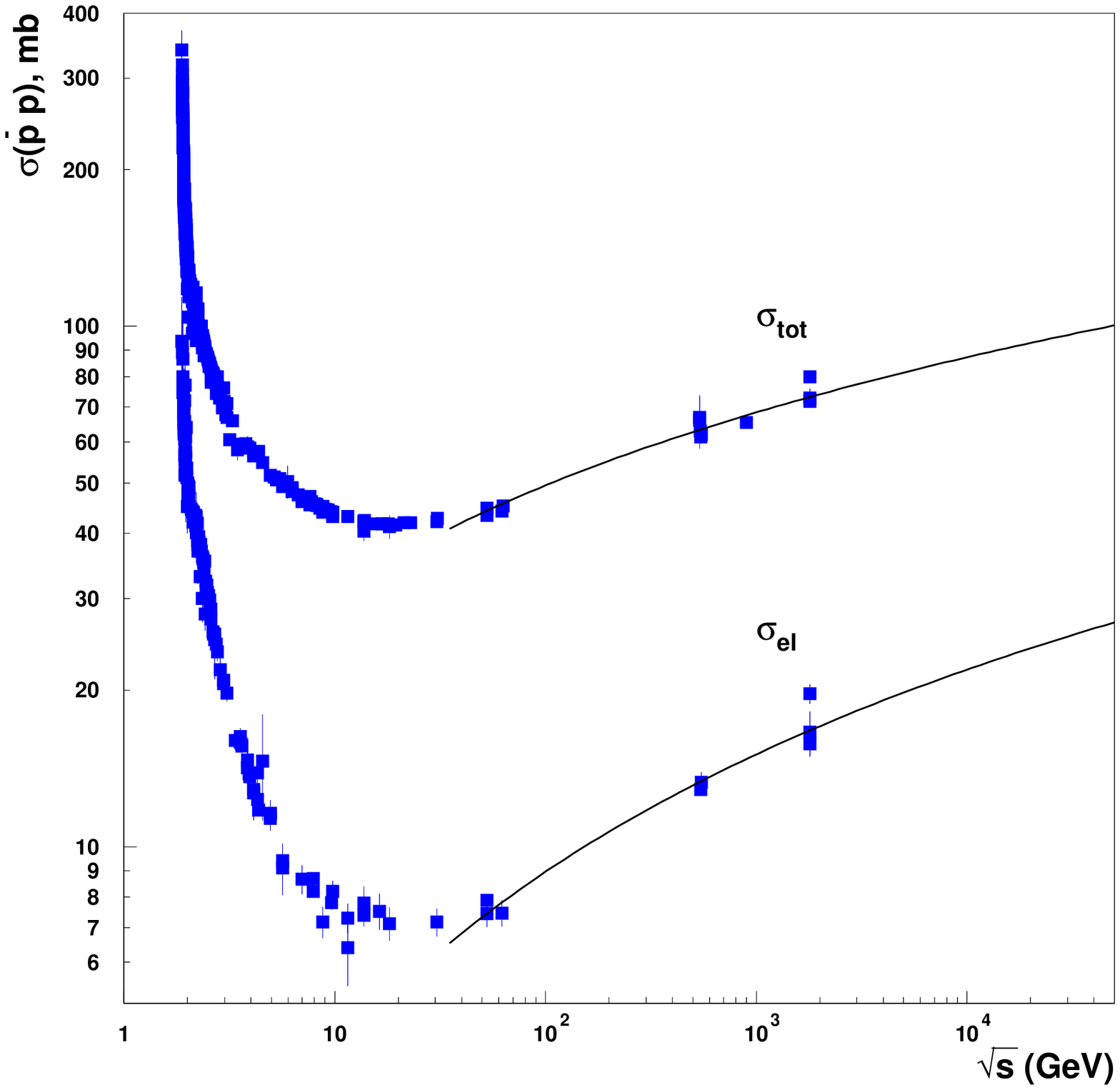}}

\parbox[t]{7.cm}{\caption{Differential cross section of
elastic $\bar pp$ scattering at the energies
$\sqrt{s}=546$ GeV and $1.8$ TeV. 
\label{fig:pbarp}}}
\hfill~\parbox[t]{7.cm}{\caption{Elastic and total cross sections of
$\bar pp$ scattering.
\label{fig:pbarpel}}}
\end{figure}

In order to use the asymptotic
formula, we choose the data on the differential cross section at energies
$\sqrt{s}=546$ GeV and $1.8$ TeV \cite{CDF}. As we take into account 
neither Pomeron daughters nor possible odderon contributions,
we concentrate on the region of low $|t|$, $0<|t|<0.2 \; {\rm GeV^2}$. The
result is presented in Table~\ref{Table 3.}, with  
$\chi^2/{\rm d.o.f.} = 1.04 $.
{\small
\begin{table}[H]
\begin{center}
\begin{tabular}{ll}
a = & $0.41 \pm 0.01\; [{\rm GeV}^{-1}]$, \\
b = & $7.61   \pm    3.36\; [{\rm GeV}^{-2}]$, \\
c = & $-1.12 \pm 1.40 [{\rm GeV}^{-3}]$, \\
d = & $7.72\pm 0.52\; [{\rm GeV}^{-2}]$, \\
g = & $2.86 \pm 0.41 $. \\
\end{tabular}
\end{center}
\vskip -0.5cm \caption{ Values of parameters obtained by fitting
$p\bar p$ data.   
\label{Table 3.}}
\end{table}}
In Figs.~\ref{fig:pbarp},~\ref{fig:pbarpel} we depict respectively the 
results of the fit for the differential cross section and the predicted
total and elastic cross sections of $\bar pp$ scattering.

The non linear trajectory of Eq.~(\ref{d3}) provides a satisfactory agreement
with the data also for this hadronic process.
We consider the obtained result as an argument in support of the Pomeron
universality.

\section{Conclusions}

The aim of this paper was to study the Pomeron exchange in reactions where
non leading contributions are absent or negligible. We have chosen $J/\Psi$
and $\Phi$ photoproduction and electroproduction as Pomeron filters.

Our analysis is based on the dipole Pomeron model assuming a Pomeron trajectory with 
intercept equal to one and a non linear $t$-dependence. The choice of the 
vertices is based on covariant Reggeization as explained in Section 2 of
Ref. \cite{FP}. To reduce the number of free parameters we have used 
an approximate form of the vertex. As a result, we have obtained a 
good description of the data on $J/\Psi$ and $\Phi$ photoproduction and electroproduction.

To demonstrate the universality of the chosen trajectory we applied the model
to $\bar pp$ scattering at sufficiently high energies where only the Pomeron
contributes.The good agreement with the experimental data is an argument 
in favor of the chosen Pomeron trajectory.

We are convinced to have reached a deeper understanding of the properties of
the dipole Pomeron.

\section*{Acknowledgments}

We would like to thank Alessandro Papa
for helpful comments and Michele Arneodo and Alessia Bruni for fruitful discussions 
on the ZEUS data. L.J. is grateful to the Department of Theoretical
Phy\-sics of the University of Torino and the Departments of Physics of 
the Universities of Calabria and Padova, together with the Istituto 
Nazionale di Fisica Nucleare - Sezioni di Padova and Torino and Gruppo 
Collegato di Cosenza, where this work was done, for their warm 
hospitality and financial support.

\vfill \eject

\end{document}